\documentclass[%
 reprint,
 amsmath,amssymb,
 aps,
]{revtex4-2}

\usepackage{graphicx}% Include figure files
\usepackage{dcolumn}% Align table columns on decimal point
\usepackage{bm}% bold math
\usepackage{xcolor}
\usepackage{times}
\DeclareGraphicsExtensions{.pdf,.eps,.png,.jpg,.mps} 
\usepackage{array, makecell} %table

\begin{document}

\title{Chip-scale high-performance photonic microwave oscillator}

\author{Yang He$^{1}$, Long Cheng$^{1}$, Heming Wang$^{1}$, Yu Zhang$^{2}$, Roy Meade$^{2}$, \\Kerry Vahala$^{3}$, Mian Zhang$^{2}$, Jiang Li$^{1}\dagger$\\
$^1$hQphotonics Inc, 2500 E Colorado Blvd Suite 330, Pasadena CA, 91107, USA\\
$^2$HyperLight Corporation, 1 Bow Street, Suite 420, Cambridge, MA 02138, USA\\
$^3$T. J. Watson Laboratory of Applied Physics, California Institute of Technology, Pasadena, CA 91125, USA\\
$^\dagger$Corresponding authors: jiang.li@hqphotonics.net}

%\date{\today}% 

\begin{abstract}
Optical frequency division \textcolor{black}{based on bulk or fiber optics} provides unprecedented spectral purity for microwave oscillators. \textcolor{black}{To extend the applications of this approach,  the big challenges are to develop miniaturized optical frequency division oscillators without trading off phase noise performance.} In this paper, we report a chip-scale electro-optical frequency division  microwave oscillator with ultra-low phase noise performance. Dual laser sources are co-self-injection-locked  to a single silicon nitride spiral resonator to provide a record high-stability, \textcolor{black}{fully} on-chip optical reference. 
\textcolor{black}{An integrated electro-optical frequency comb based on  a novel thin-film lithium niobate   phase modulator chip is incorporated for the first time to perform optical-to-microwave frequency division}.   The resulting chip-scale photonic microwave oscillator  achieves a phase noise level of -129 dBc/Hz at 10 kHz offset for 37.7 GHz carrier.  The results represent a major advance in high performance, integrated photonic microwave oscillators for applications including signal processing, radar, timing, and coherent communications.
\end{abstract}

\maketitle 

\section*{Introduction}

Optical frequency division (OFD) \cite{Diddams2020} is the preeminent approach for generation of high-performance  microwave signals \cite{fortier2011generation}, offering the lowest phase noise microwave signals at X band (around 12 GHz) \cite{xie2017photonic} \textcolor{black}{ based on cavity-stabilized lasers and  fiber-based frequency combs.}  A variation of OFD   called electro-optical frequency division (eOFD) performs optical-to-microwave frequency division using electro-optical frequency combs \cite{li2014electro}.  Compact  turn-key eOFD oscillators with record low phase noise at K band (40 GHz) have been \textcolor{black}{ demonstrated \cite{Li:23}, reducing the form factor of high performance OFD systems to less than 3 liters including all optics and electronics.  To expand the application reach of these signal sources in mobile, airborne, and chip-scale systems, the development of miniaturized OFD oscillators that reduce size, weight and power without trading off phase noise performance is of keen interest.} And recently, \textcolor{black}{soliton microcombs} have been used to achieve impressive low phase noise levels by OFD \cite{kudelin2023photonic,sun2023integrated}. Besides OFD and eOFD systems, other types of photonic microwave oscillators include microcombs \cite{kippenberg2018dissipative,li2012low, yi2015soliton,matsko2016turn,kwon2022ultrastable,weng2019spectral,lucas2020ultralow,yao2022soliton,yang2021dispersive}, mode-locked laser frequency combs \cite{jung2013ultralow,kalubovilage2022x}, optoelectronic oscillators \cite{maleki2011optoelectronic}, and on-chip Brillouin oscillators \cite{li2013microwave,li2014low,gundavarapu2019sub,merklein2016widely}, \textcolor{black}{among which the oscillator phase noise levels  typically need  to trade off with form factor and power. }

\textcolor{black}{Integration of the OFD or eOFD based microwave oscillators} requires integration of two key elements: \textcolor{black}{III-V lasers with  optical reference cavities}, and optical frequency combs. Considering reference cavities first, \textcolor{black}{spiral resonators \cite{Lee:2013,BLi:21,liu202236}} make excellent \textcolor{black}{on-chip} frequency references. These resonators suppress thermorefractive noise (TRN) through their large mode volumes \cite{Lee:2013}. Moreover, self-injection-locking of a single DFB laser to a Si$_3$N$_4$-based spiral reference dramatically suppresses TRN noise while also improving short-term coherence \cite{BLi:21}. Pound-Drever-Hall (PDH) frequency locking of dual external-cavity lasers to on-chip disk microresonators \cite{li2014electro} and spiral resonators \cite{sun2023integrated} has also been shown to provide dual laser references with additional stability from common-mode rejection.

\textcolor{black}{Concerning integrated frequency combs, alongside soliton microcombs \cite{kippenberg2018dissipative} another type of integrated microcomb, the thin-film LiNbO$_3$ (TFLN) electro-optic comb (EO comb), has also witnessed rapid progresses in recent years. TFLN is offering remarkable advances in low-V$_\pi$ phase and intensity modulators \cite{wang2018integrated,kharel2021breaking,yu2022integrated,xu2022dual,ahmed2020high,xue2023full,valdez2023100}. TFLN EO combs have been generated in ring resonators \cite{zhang2019broadband, hu2022high}, as well as using non-resonant phase modulators and amplitude modulators \cite{ren2019integrated, yu2022integrated, zhang2023power}. They also feature tunable repetition rates and rapid tuning speeds, which are desirable in OFD systems for low phase noise generation within a large servo locking bandwidth.}

\begin{figure*}[t]
\centering
\includegraphics[width=\linewidth]{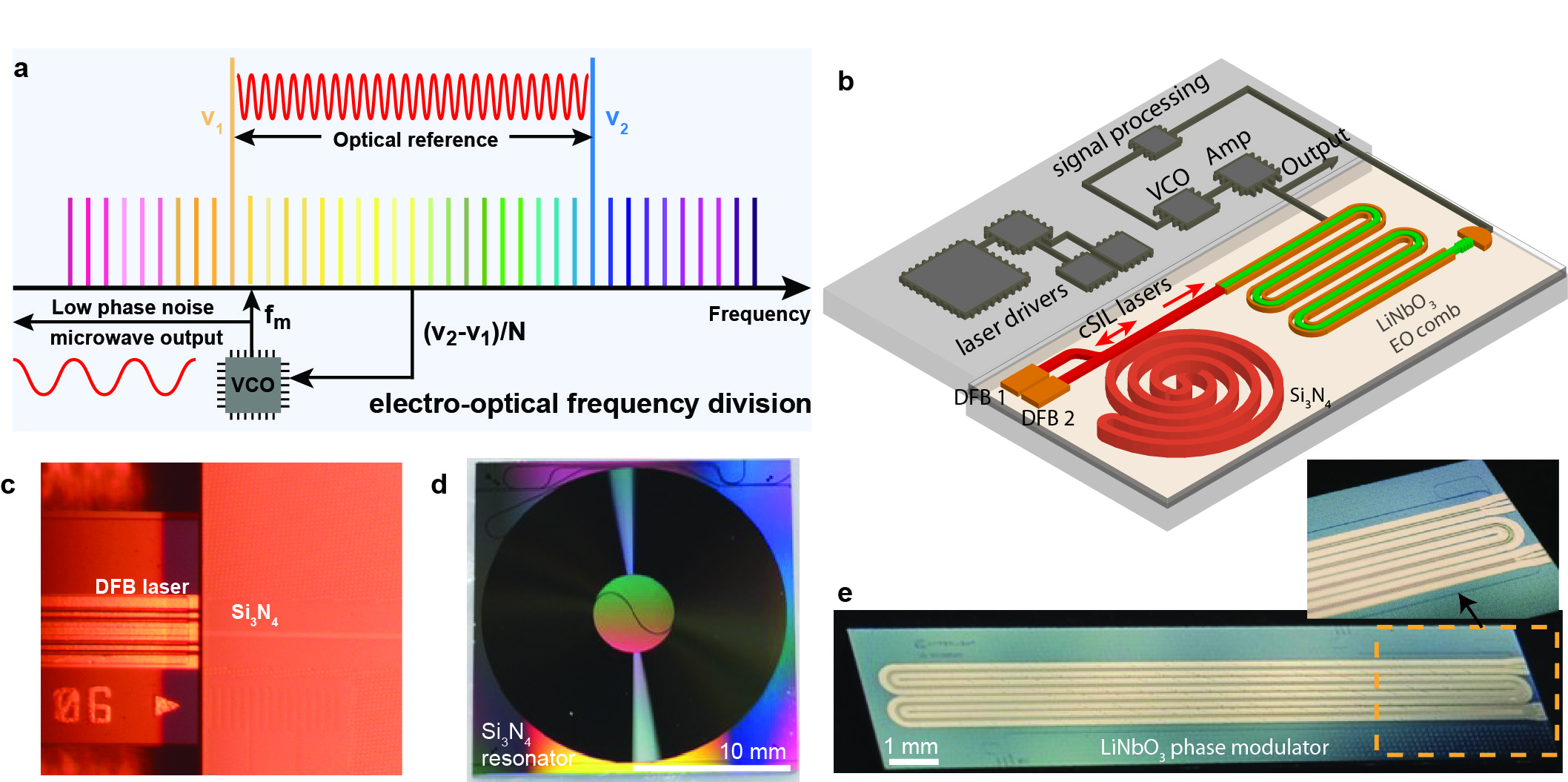}
\caption{ \textbf{(a)} Principle  of electro-optical frequency division (eOFD) is shown. Dual laser reference ($\nu_1$ and $\nu_2$)  are strongly phase modulated (LiNbO$_3$ modulator) to generate two electro-optic (EO) combs. Phase noise from the the voltage controlled oscillator (VCO) used to drive the LiNbO$_3$ modulators is transferred into the line spacing noise of the two EO combs. At the spectral mid point between $\nu_1$ and $\nu_2$, the far upper sideband of $\nu_1$ nearly overlaps with the far lower sideband of $\nu_2$. A beat note $\delta f$ is generated between these two nearly overlapping EO comb lines, which is given by $\delta f =  |\nu_2 - \nu_1 - Nf_m|$, where N is the number of EO comb lines between $\nu_1$ and $\nu_2$. This beat note contains the free-running phase jitter of the microwave VCO multiplied by a factor of N. When this jitter is nulled by servo control of the VCO tuning port, the closed loop VCO modulation frequency is given by $f_m = |\nu_2 - \nu_1 \pm \delta f|/N$, and is locked to a small fraction (1/N) of the optical reference ($\nu_2 - \nu_1$). The phase noise of the VCO is divided by a factor of $N^2$ relative to the dual optical references (within the servo lock bandwidth) \cite{li2014electro}. \textbf{(b)} Schematic concept design for a chip-scale integrated eOFD oscillator based on a fully on-chip dual laser reference (co-self-injection-locked (cSIL) lasers to an ultra-high Q spiral resonator) and a thin-film LiNbO$_3$ (TFLN)  EO comb.   \textbf{(c)} Microscope image of DFB coupling to one input waveguide of the spiral resonator. In the complete cSIL devices a second DFB laser is coupled to another Si$_3$N$_4$ input waveguide. \textbf{(d)} Photograph of the 14m-long Si$_3$N$_4$ spiral resonator with footprint 21$\times$21mm. Scale bar at lower right corner: 10 mm. \textbf{(e)} Photograph of a low V$_\pi$  thin-film LiNbO$_3$ (TFLN) phase modulator chip with recycled design (double pass) and bent electrodes.  Chip footprint is 13.75 mm$\times$3.5 mm.  Scale bar at lower left corner: 1 mm.  Zoom-in view of the right side of the TFLN chip is also shown.}
\label{fig1}
\end{figure*}

In this work, we report critical advances in chip-scale  photonic microwave oscillators. A  chip-scale high performance photonic microwave oscillator is developed based on a fully on-chip dual laser reference and an integrated TFLN EO comb. First, the low phase noise dual laser reference  \textcolor{black}{(including two III-V lasers and on-chip reference cavity)} is demonstrated based on co-self-injection-locking (cSIL) of two DFB laser chips to a single silicon nitride spiral resonator. This dual reference achieves a record-low on-chip optical phase noise and greatly simplifies the OFD system architecture. Specifically, the cSIL reference obviates the need for  off-chip components such as optical isolators, external-cavity lasers and multiple laser frequency locking components, as used in other OFD systems.   Second, \textcolor{black}{an integrated EO comb is utilized for the first time to perform optical-to-microwave frequency division. The integrated EO comb is generated by pure phase modulation from a thin-film lithium niobate  phase modulator chip.} The resulting integrated eOFD oscillator achieves a phase noise of -129 dBc/Hz at 10 kHz offset for 37.7 GHz carrier output, which is equivalent to -141 dBc/Hz at 10 kHz offset for a 10 GHz carrier. This matches the record low phase noise recently reported for a photonic chip based, soliton OFD microwave oscillator   \cite{kudelin2023photonic}.

\section*{System Design}

A brief description of the principle of eOFD and the schematic concept design for the integrated eOFD oscillator are provided in the Figures \ref{fig1}a and \ref{fig1}b, respectively. Referring to Figure \ref{fig1}b, the signal source uses dual DFB lasers that are co-self-injection locked to an ultra-high-Q Si$_3$N$_4$ spiral reference cavity. Besides system simplifications noted above compared with PDH-locked frequency stabilized lasers, the cSIL approach has the added benefit that relative coherence of the two locked lasers is inherently insensitive to common mode frequency noise of the Si$_3$N$_4$ spiral cavity. A photograph of the Si$_3$N$_4$ spiral resonator along with endfire coupling to one of the DFB lasers is shown in Figures \ref{fig1}c. The chip-scale  spiral resonator (shown in Figure \ref{fig1}d with a footprint of 21mm $\times$ 21mm) has a 14 m round-trip length for TRN suppression while achieving a record-high spiral resonator Q factor of 332 million.

The integrated electro-optic comb is generated by phase modulation with a novel phase modulator TFLN chip, shown in Figure \ref{fig1}e. The TFLN phase modulator chip incorporates recycled optical waveguides, which double the phase modulator length given a fixed length of co-planar waveguide electrodes. The bent-electrode design increases total electrode length to 50 mm in a small foot print of only 13.75 mm $\times$ 3.5 mm, as shown in Figure \ref{fig1}e. Significantly, the recycled optical waveguide design reduces V$_\pi$ and drive power for broadband EO comb generation. \textcolor{black}{Moreover, not only does the TFLN EO comb frequency divider remove an additional pump laser   for soliton microcomb generation, but it also obviates   highly linear ultrafast photodetectors required in soliton OFD systems, further reducing system complexity \cite{li2014electro}.}

\begin{figure*}[t]
\centering
\includegraphics[width=\linewidth]{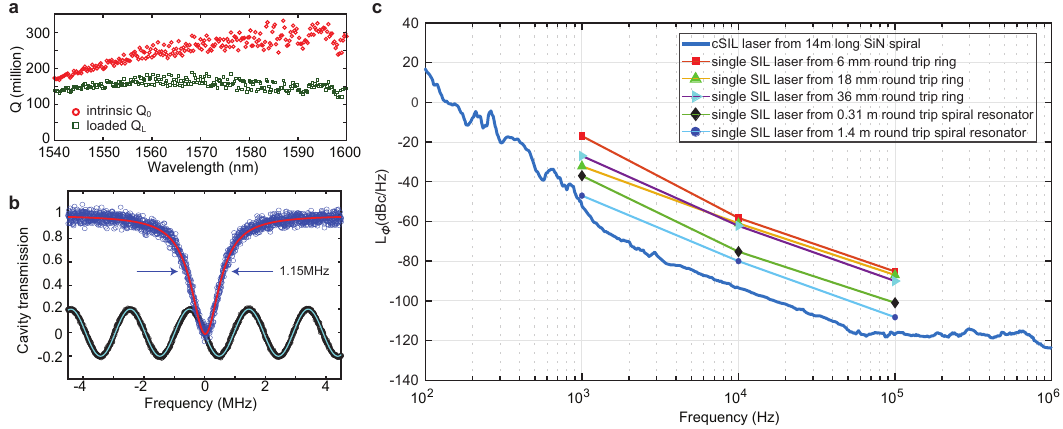}
\caption{ \textbf{(a)} Measured intrinsic (Q$_0$, red circle markers) and loaded   (Q$_L$, green square markers) Q factors for the 14m-long Si$_3$N$_4$ spiral resonator are plotted versus wavelength from 1540 nm to 1600 nm.  \textbf{(b)} Cavity transmission of the ultra-high-Q mode at 1587 nm in the 14m-long Si$_3$N$_4$ spiral resonator shows a linewidth of 1.15 MHz, loaded Q$_L$ of 166 million, and intrinsic Q$_0$  of 332 million. The sine wave is the Mach-Zehder interferometer output for laser frequency calibration. \textbf{(c)} Phase noise of dual DFB lasers co-self-injection-locked (cSIL) to the ultra-high-Q, 14m-long, Si$_3$N$_4$ spiral resonator (blue curve). The phase noise is independent of the frequency separation of the dual cSIL lasers and is -94 dBc/Hz at 10 kHz offset. For comparison, the phase noise spectra of a single laser self-injection-locked to Si$_3$N$_4$ ring or spiral resonators with different round trip lengths (6 mm, 18 mm, 36 mm, 0.35 m, 1.4 m) are shown \cite{BLi:21, jin2021hertz}.}
\label{fig2}
\end{figure*}

\section*{Spiral resonator cSIL operation}

The ultra-high-Q Si$_3$N$_4$ spiral resonator  is based on the ultra-low loss, low confinement Si$_3$N$_4$ waveguides fabricated at a CMOS foundry \cite{BLi:21,jin2021hertz}.  It consists of interleaved (inward and outward) Archimedean spiral waveguides. An S-shaped waveguide connects the interleaved spirals with a rotation symmetry of 180 degree.  Limited by the maximum reticle size of 21 mm x 21 mm of the stepper photolithography tool, a maximum round-trip length of 14 meters is achieved (see Figure 1d).  The loaded (green markers) and intrinsic Q factors (red markers) measured over a wavelength span of 60 nm encompassing C-band are shown in Figure 2a. Figure 2b shows the Lorentzian fitting for the cavity transmission at 1587 nm, which features a record-high intrinsic Q factor of 332 million for spiral resonators. A summary of the measured Q factors from other spiral resonators is provided in Table S1 of the Supplementary Information. 

To characterize the relative phase noise of DFB lasers under cSIL operation, two DFB lasers with center frequency separation of only 20 GHz were co-self-injection-locked to the 14m long Si$_3$N$_4$ spiral resonator. Their beat note was then detected on a fast photodetector with bandwidth of 40 GHz, and the phase noise of the 20 GHz beat note was measured (blue curve in Figure 2d). A relative phase noise of -94 dBc/Hz at 10 kHz offset is measured. For comparison, the absolute phase noise of a single laser self-injection-locked to ultra-high-Q Si$_3$N$_4$ ring resonators and Si$_3$N$_4$ spiral resonators having different round-trip lengths is shown in Figure 2c (data from \cite{BLi:21,jin2021hertz}). Consistent with the scaling of TRN noise observed in these previous measurements, the cSIL laser phase noise (from 1 kHz to 100 kHz offsets)  scales approximately inversely with the cavity volume. Specifically, the 14m long Si$_3$N$_4$ spiral resonator phase noise (-94 dBc/Hz at 10 kHz offset) is suppressed by over 10 dB relative to the phase noise (-80 dBc/Hz at 10 kHz offset) of the 1.4m long Si$_3$N$_4$ spiral resonator in \cite{BLi:21}. Since the cSIL noise results from the beat noise of two lasers, it is expected to be 2X larger than the underlying noise of each laser so that the overall noise is about 17 dB lower in the 14m cSIL device. The white phase noise floor above 60 kHz offset is due to the white thermal noise from the fast photodetector. Comparing the dual locked laser phase demonstrated here to earlier reports (all at 10 kHz offset), dual on-chip Brillouin laser references have a phase noise of -90 dBc/Hz using a silica disk resonator \cite{li2013microwave}, and -84 dBc/Hz using a Si$_3$N$_4$ ring resonator \cite{gundavarapu2019sub} at 10 kHz offset. And dual laser references by Pound-Drever-Hall frequency locking of two external-cavity lasers to a Si$_3$N$_4$ spiral resonator have achieved a phase noise of -80 dBc/Hz \cite{sun2023integrated}.

\begin{figure*}[t]
\centering
\includegraphics[width=\linewidth]{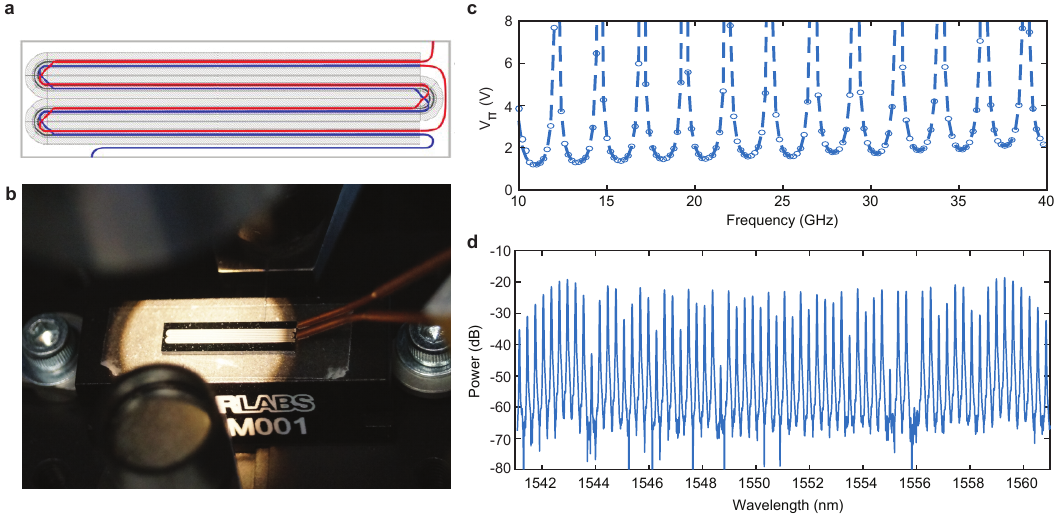}
\caption{ \textbf{(a)} Schematic of the recycled TFLN phase modulator chip with bent electrodes. Total chip size is 13.75 mm x 3.5 mm. Total length of electrode (ground-signal-ground or GSG type) is 50 mm , with three U bends and four straight sections. The TFLN optical waveguides are based on a recycled, two-pass design. The blue (red) trace represents the first (second) pass among the ground-signal electrodes. \textbf{(b)} Photograph of the TFLN phase modulator RF and optical alignment set up. Lensed fibers are used for optical input and output coupling. A dual RF probe (each with GSG configuration) is used for RF input and output coupling. \textbf{(c)} Measured V$_\pi$ of the recycled TFLN phase modulator chip with respect to modulation frequency is shown. \textbf{(d)} Generated broadband EO comb spectrum (38 GHz line spacing, 3dB bandwidth of 2.2 THz) from the TFLN recycled phase modulator chip is shown.  } 
\label{fig3}
\end{figure*}

\begin{figure*}[t]
\centering
\includegraphics[width=0.9\linewidth]{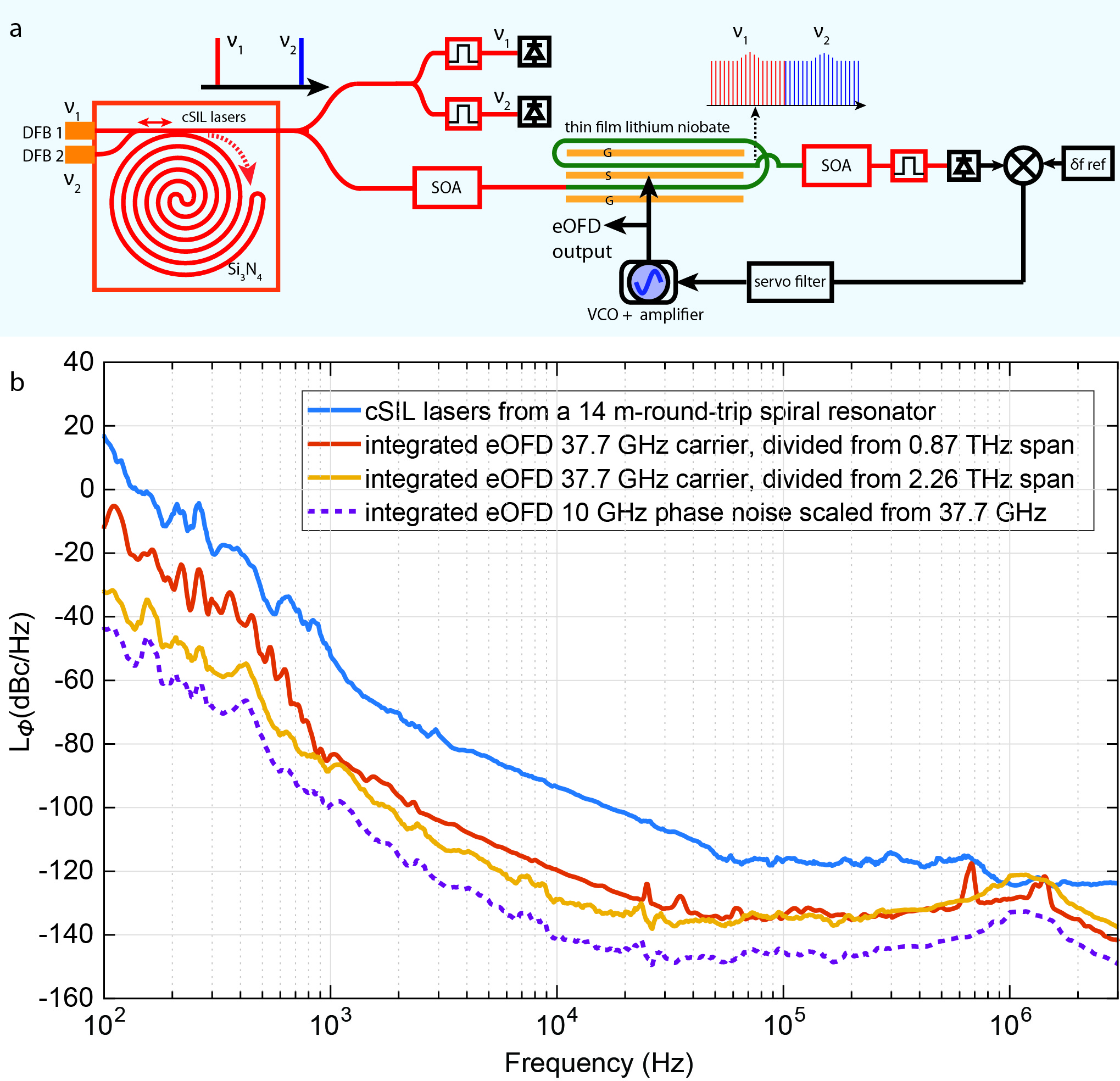}
\caption{\textbf{(a)} Experimental setup for the chip-scale eOFD oscillator. Two DFB lasers are co-self-injection-locked (cSIL) to an ultra-high-Q Si$_3$N$_4$ spiral resonator.  A 50/50 Si$_3$N$_4$ waveguide directional coupler combines the two laser inputs before coupling to the spiral resonator. The cSIL laser references  are then coupled out at the Si$_3$N$_4$ through port   and amplified by a semiconductor optical amplifier (SOA). After the SOA, the cSIL lasers are coupled to the TFLN low-V$_\pi$ phase modulator chip using a lensed fiber to generate on-chip EO combs with line spacing of 37.7 GHz.  The spectral middle point lines between dual reference lasers are optically bandpass filtered and detected for amplified phase error detection of the EO comb line spacing phase noise. The eOFD phase error is then used for feedback control of the VCO via a fast servo filter.
\textbf{(b)} Phase noise of the chip-scale eOFD oscillator is shown. The blue curve is the reference phase nosie of the cSIL dual laser reference from the 14m long Si$_3$N$_4$ spiral resonator. The red (yellow) curve is the phase noise of the chip-scale eOFD oscillator at 37.7 GHz carrier, divided from 0.87 (2.26) THz cSIL dual laser span. A phase noise level of -129 dBc/Hz at 10 kHz offset, \textcolor{black}{ and -136 dBc/Hz at 50 kHz offset} with microwave power of 17 dBm was obtained for the 37.7 GHz carrier (yellow curve). The dashed purple curve is the phase noise
scaled to a 10 GHz carrier frequency from the 37.7 GHz output (vertical scale offset by 20Log$_{10}$[37.7 GHz/10 GHz]), showing an equivalent phase noise of -141 dBc/Hz at 10 kHz offset, which is an equal-record low phase noise with the recent demonstration of a soliton-based OFD  system \cite{kudelin2023photonic}.  }
\label{fig4}
\end{figure*}

\section*{Integrated  LiNbO$_{3}$ EO comb generation}

A single TFLN phase modulator with low V$_\pi$ performance was developed in this work for broadband integrated electro-optic comb generation. Figure 3a gives details on the layout of the TFLN modulator.  First, an electrode length of 50 mm was used, wherein the electrodes bend three times to create four rows of straight electrodes. Second, a recycled design for the TFLN optical waveguides was implemented to double the overall electro-optic modulation length to 100 mm. The blue (red) trace in Figure 3a shows the first (second) optical pass among the electrodes, and two passes (red and blue traces) are connected via a U bend on the right side of the TFLN chip.  \textcolor{black} {A similar recycled design for the TFLN phase modulator was implemented previously using straight electrodes to double the EO modulation length \cite{yu2022integrated}}. These features reduce V$_\pi$ while also maintaining a compact modulator footprint of 13.75 mm $\times$ 3.5 mm. Micro-structured electrodes were also implemented in this device to improve the modulator frequency response at higher frequencies similar to those demonstrated in \cite{kharel2021breaking}. 

Figure 3b shows the optical and RF coupling setup for the phase modulator. Two lensed
fibers were used to couple light into and out of the TFLN chip. And a dual RF probe with ground-signal-ground configuration was used to launch microwave signals into the input electrode, and terminate the RF electrode end at the output electrode. The measured V$_\pi$ of the modulator plotted versus modulation frequency is shown in Figure 3c. Due to the recycled design, the in-phase superposition of the phase modulation between the first pass and the second pass causes a 2X reduction of the V$_\pi$ at frequencies (N+1/2)FSR, where FSR is the free spectral range and N is an integer.  The measured V$_\pi$ is 1.5 V at 18 GHz, and 2.1 V at 40 GHz. To the best of the our knowledge, these are record-low V$_\pi$ values for a LiNbO$_{3}$ phase modulator at these frequencies for telecomm C-band wavelengths. For comparison, the two-pass, recycled TFLN phase modulator in \cite{yu2022integrated} has a V$_\pi$ of 2.6 V at 18.5 GHz. Also, a quad-pass recycled TFLN phase modulator has a V$_\pi$ of 2.5 V at 20 GHz \cite{zhang2023power}. Table S2 in the Supplementary Information summarizes V$_\pi$ measurements from various LiNbO$_{3}$ phase modulators at C band wavelengths. Finally, the resulting electro-optic comb with 38 GHz line spacing and 2.2 THz bandwidth (3 dB) (launched RF power of 36 dBm) is shown in Figure 3d.

\section*{Oscillator noise performance}

Figure \ref{fig4}a shows the experimental setup for the chip-scale eOFD oscillator.  Two cSIL DFB lasers  at wavelengths 1542 nm and 1559 nm are used to create a  2.26 THz frequency span optical reference. A 50/50 Si$_3$N$_4$ waveguide directional coupler combines the two input ports before coupling to the spiral resonator using a pulley coupler design. The cSIL laser signals are then coupled out at the Si$_3$N$_4$ through port ($\sim$ 2 mW) and amplified by a semiconductor optical amplifier (SOA) to  25 mW. After the SOA, the cSIL signals are coupled to the TFLN low-V$_\pi$ phase modulator chip using a lensed fiber. The resulting EO combs  (with line spacing 37.7 GHz) overlap at the spectral middle point between the cSIL laser lines. The EO combs are post-amplified using a second SOA to 5 mW. The spectral middle point lines between dual reference lasers are optically bandpass filtered and detected. The eOFD phase error is then generated and used for feedback control of the VCO via a fast servo filter. The optical-to-microwave frequency division factor of  N = 60 results in a phase noise reduction of 35 dB from the optical reference to the microwave carrier at 37.7 GHz. The measured phase noise spectrum of the 37.7 GHz carrier is presented in Figure 4b (yellow curve). A phase noise level of -129 dBc/Hz at 10 kHz offset, \textcolor{black}{ and -136 dBc/Hz at 50 kHz offset} with microwave power of 17 dBm is  obtained for the 37.7 GHz carrier.  

For comparison, the phase noise of the cSIL optical reference signal is also shown in the figure (blue curve) as well a second electro-optical frequency division measurement using cSIL lasers with a reduced frequency span of 0.87 THz (red curve, division factor of 23 to 37.7 GHz). The dashed purple curve in Figure 4b is the phase noise scaled to a 10 GHz carrier frequency (vertical scale offset by 20 Log$_{10}$[37.7 GHz/10 GHz]), showing an equivalent phase noise of -141 dBc/Hz at 10 kHz offset, \textcolor{black}{ and -148 dBc/Hz at 50 kHz offset.} The 10 kHz offset phase noise level of -141 dBc/Hz for 10 GHz carrier is a record-low phase noise for chip-scale photonic microwave oscillators; and on par with a recent demonstration using a photonic-chip-based, soliton microcomb OFD system \cite{kudelin2023photonic}. \textcolor{black}{ Finally, the closed-loop integrated eOFD servo locking bandwidth is $>$1 MHz, which is $>$3X higher than the servo locking bandwidth of  current soliton-OFD systems \cite{kudelin2023photonic,sun2023integrated}. }

\section*{Discussion and Conclusion}

\textcolor{black}{The core components of this oscillator demonstration include four types of photonic chips: two DFB lasers, Si$_3$N$_4$ spiral resonator, TFLN modulator, and InGaAs photodetector. Hybrid integration \cite{spencer2018optical}, or heterogeneous integration of III-V with Si$_3$N$_4$ \cite{xiang2021high,xiang2023three} and Si$_3$N$_4$ with TFLN \cite{chang2017heterogeneous,churaev2023heterogeneously} can be leveraged for scalable production of the oscillator. The two semiconductor optical amplifiers used in the current demonstration can be omitted in future designs by increasing cSIL laser power, and reducing the Si$_3$N$_4$ to TFLN coupling loss and the TFLN on-chip propagation loss. Moreover, the current integrated eOFD demonstration used a 2.26 THz frequency span reference, limited by the non-resonant TFLN EO comb optical bandwidth.  Along these lines, recent EO combs based on TFLN ring resonators produce much broader EO comb bandwidths \cite{zhang2019broadband,hu2022high}.} \textcolor{black}{ Integrated eOFD with wider resonant EO combs  would lead to additional $>$10 dB phase noise reductions by further increasing the optical-to-microwave division ratio. }

In conclusion, a chip-scale photonic microwave oscillator with record-low phase noise has been reported. The oscillator implements integrated electro-optical frequency division  with two significant chip-scale device innovations. First, a fully on-chip, dual optical reference is demonstrated based on a monolithic ultra-high-Q silicon nitride spiral resonator operated in co-self-injection-locking mode with two DFB lasers. This cSIL demonstration dramatically simplifies the frequency reference by eliminating various off-chip components such as optical isolators, external-cavity lasers, and laser frequency locking components required in other OFD systems. Moreover, it provides \textcolor{black}{record low phase noise for on-chip lasers } through the record high-Q and large-mode-volume spiral design. Second, an integrated low-V$_\pi$ phase modulator based on a recycled, dual pass geometry design enables wide EO comb bandwidth at reduced drive power. The simplified architecture eliminates several subsystems and components that are typically required in OFD systems, thereby reducing complexity and ultimate system size. \textcolor{black}{Also, the mass-producible, solid-state optical reference demonstrated here improves robustness and manufacturing economy of scale. }  \textcolor{black}{ The high performance, chip-scale photonic microwave oscillator demonstrated here represents a major advance in integrated photonic microwave oscillators and is expected to have major performance impacts on many applications including precision timing, signal processing, radar and coherent communications.  }

\bibliography{main.bib}% Produces the bibliography via BibTeX.

\medskip
\medskip

\newpage

\newpage 

\section*{Methods}

The Si$_3$N$_4$ spiral resonator is based on the low-confinement, high aspect ratio  Si$_3$N$_4$ waveguides, fabricated  in a COMS-ready foundry \cite{BLi:21,jin2021hertz}.  The detailed fabrication processes are described in \cite{jin2021hertz}. The Si$_3$N$_4$ core is cladded by thermal SiO$_2$ (lower cladding), and   LPCVD SiO$_2$ (upper cladding). An on-chip Si$_3$N$_4$ directional coupler with 50/50 coupling ratio is placed on the input bus waveguide, to combine the two DFB laser inputs before the Si$_3$N$_4$ spiral  resonator. 

For Si$_3$N$_4$ spiral resonator Q measurement, a  continuously-tuning, external-cavity diode laser (Toptica CTL 1550) was used to scan across the resonator resonance frequencies. The instantaneous, short term linewidth of the scanning laser is $<$ 10 kHz, much smaller than the resonator linewidth ($\sim$ 1 MHz) in our measurements. The frequency scan range of the laser is calibrated and measured at the same time using a separate Mach-Zehnder Interferometer (MZI) while the laser is scanning across the resonator resonances. 

For TFLN device fabrication, commercial x-cut LN-on-insulator wafer (NANOLN) is used. On the wafer, a thin-film LiNbO$_3$ layer stays on top of a SiO$_2$/Si stack substrate. Deep UV (DUV) and Ar+ based reactive ion etching are used to define the optical waveguides in thin film LiNbO$_3$. The entire device is cladded with silicon dioxide via plasma-enhanced chemical vapor deposition. Then metal electrodes are patterned using a self-aligning lift-off process. In the end, the chip edges are diced for fiber to chip coupling.   
 
For V$_\pi$ measurement of the TFLN phase modulator chip, a high resolution optical spectrum analyzer  was used to measure the first order EO sideband power relative to zeroth order laser power under EO phase modulation, from which the phase modulation depth ($\delta \phi = \pi V/V_{\pi}$ ) and the modulator $V_{\pi}$    were calculated at each microwave modulation frequency ($f_M$) according to the Jacobi–Anger expansion of phase modulated electrical fields:
\begin{equation}
E_0 e^{i\omega_0 t} e^{i\delta \phi  cos(2\pi f_M t) } = E_0 e^{i\omega_0 t}  \sum_{n=-\infty}^{+\infty} i^n J_n(\delta \phi)e^{in2\pi f_M t}
\label{eq1}
\end{equation} 
 
where $\omega_0$ is the laser carrier angular frequency, $J_n(\delta \phi)$ is the n-th Bessel function of the first kind. 

A lensed fiber was used to couple light out from the Si$_3$N$_4$ spiral resonator chip with an edge coupling loss of 2 dB. Two lensed fibers were used to couple light into and out from the TFLN chip with a fiber-to-fiber insertion loss of 13 dB.  The lensed fiber to TFLN edge coupling loss is measured at 4 dB/facet. Therefore, the  on-chip propagation loss for the TFLN phase modulator is 5 dB.

For the phase noise measurement, an ultra-low phase noise 40 GHz eOFD oscillator with a compact modular form factor described in \cite{Li:23} was used to down-convert the 37.7 GHz chip-eOFD oscillator signal to 2.3 GHz. The phase noise of the 2.3 GHz signal was then measured by a commercial phase noise analyzer (Rohde Schwarz FSUP26). The phase noise of the reference 40 GHz eOFD oscillator (-153 dBc/Hz at 10kHz offset for 40 GHz carrier) is much lower than the phase noise of chip-scale eOFD oscillator.

\medskip

\noindent\textbf{Funding.}  This material is based upon work supported by the Defense Advanced Research Projects Agency (DARPA) GRYPHON program under Contract No. HR001122C0019.
\medskip
 
\noindent\textbf{Author Contributions.} The concepts were conceived by J.L. and K.V.  Measurements and analysis were performed by Y.H. and L.C. The Si$_3$N$_4$ spiral resonator was designed by H.W. and J.L. The TFLN modulator was designed and fabricated by Y.Z., R.M. and M.Z. All authors contributed to the writing of the paper. J.L. designed the experiment and supervised the project.
\medskip

\noindent\textbf{Disclosures.} The authors declare no conflicts of interest.
This research was developed with funding from the Defense Advanced
Research Projects Agency (DARPA). The views, opinions and/or findings expressed are those of the author and should not be interpreted as representing the official views or policies of the Department of Defense or the U.S. Government. Distribution Statement "A" (Approved for Public Release, Distribution Unlimited).
\medskip

% SI
\clearpage
\onecolumngrid
\appendix
\renewcommand{\theequation}{S\arabic{equation}}
\renewcommand{\thefigure}{S\arabic{figure}}
\renewcommand{\thetable}{S\arabic{table}}
\setcounter{figure}{0}
\setcounter{equation}{0}

\newpage

\section*{Supplementary Information}

\section{Single pass versus double pass (recyled) TFLN phase modulator design}

 \begin{figure*}[h]
\centering
\includegraphics[width=0.8\linewidth]{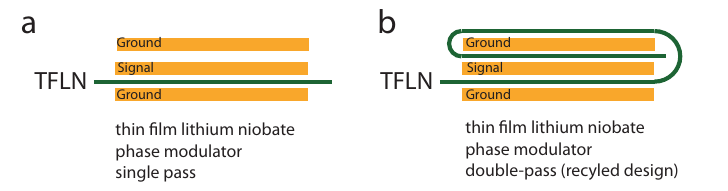}
\caption{ (a) Schematic for a single pass thin film lithium niobate phase modulator. The TFLN waveguide (green line) only passes through the gap between the signal and ground electrodes (yellow) of the coplanar-waveguide (CPW) once. (b) Schematic for a double-pass (recycled design) thin film lithium niobate phase modulator. The TFLN waveguide (green line) passes through the two gaps of the  CPW twice. }
\label{figLN}
\end{figure*}

The schematics for a single-pass and double-pass (recycled) thin film lithium niobate (TFLN) phase modulator are shown in Figure \ref{figLN}. For the single-pass design, the optical waveguide (green line) only passes through the gap between the signal and ground electrodes of the coplanar-waveguide (CPW) once. For the double-pass (recycled) design,   after the first pass through the first ground-signal gap of the CPW, the waveguide is looped back to pass through the second ground-signal gap of the CPW. 
Therefore, the EO modulation depth is doubled for the double-pass (recyled) design, and the V$_\pi$ is reduced by a factor of two, for microwave modulation frequencies at (N+1/2)FSR, where N is an integer and FSR is the free-spectral-range of the recycled phase modulator. The 1/2 term is due to the different electrical field directions for the two CPW gaps.

\section{Summary of various ultra-high-Q spiral resonators}

The summary of various ultra-high-Q (UHQ)  spiral resonators \cite{BLi:21, liu202236, Lee:2013} is given in Table \ref{tab1}, along with various UHQ Si$_3$N$_4$ ring resonators \cite{ jin2021hertz, liu2022ultralow, puckett2021422}. The Si$_3$N$_4$ spiral resonator in this work achieved an intrinsic Q factor of 332 million, which is a record high Q factor for on-chip spiral resonators. For comparison, the TE mode Si$_3$N$_4$ spiral resonator in \cite{BLi:21} has an intrinsic Q factor of 164 million, and the TM mode Si$_3$N$_4$ spiral resonator in \cite{liu202236} has an intrinsic Q factor of 80 million. The silica wedge spiral resonator has an an intrinsic Q factor of 140 million \cite{Lee:2013}.
\textcolor{black}{Note that the high-confinement Si$_3$N$_4$ ring resonators (with thicker Si$_3$N$_4$ core), with intrinsic Q factors of 30 million \cite{liu2021high}, 67 million \cite{ji2017ultra}, are not included in the table. }

\begin{table}[!htb]
    \normalsize
    \centering
    \setlength{\tabcolsep}{6mm}{
    \begin{tabular}{|c|c|c|c|}
        \hline
        \textbf{Resonator type} & \textbf{Intrinsic Q Factor Q$_0$} & \textbf{Round trip length} & \textbf{Mode}\\ 
        \hline
         SiN spiral resonator [This work]  & 332 M at 1587 nm & 14 m & TE \\ 
         \hline
         SiN spiral resonator \cite{BLi:21}  & 164 M at 1550 nm & 1.4 m & TE \\
         \hline
         SiN spiral resonator \cite{liu202236}  & 80 M at 1550 nm & 4 m & TM \\
         \hline
         SiO$_2$ spiral resonator  \cite{Lee:2013}  & 140 M at 1550 nm & 1.2 m & TE \\
         \hline
         SiN ring resonator \cite{jin2021hertz}  & 260 M at 1600 nm & 6 mm & TE \\
         \hline
         SiN ring resonator \cite{liu2022ultralow}  & 720 M at 1615 nm & 74 mm & TM \\
         \hline
         SiN ring resonator \cite{puckett2021422}  & 422 M at 1570 nm & 74 mm & TE \\
        \hline
    \end{tabular}}
    \caption{Summary of various ultra-high-Q (UHQ)  spiral resonators \cite{BLi:21, liu202236, Lee:2013}, and UHQ Si$_3$N$_4$ ring resonators \cite{ jin2021hertz, liu2022ultralow, puckett2021422}.}
    \label{tab1}
\end{table}

\begin{table}[!htb]
    \normalsize
    \centering
    
    \setlength{\tabcolsep}{7mm}{
    \begin{tabular}{|c|c|c|}
        \hline
        
        \textbf{Phase Modulator Type} & \textbf{V$_{\pi}$ } & \textbf{Architecture} \\ 
        \hline
         TFLN phase modulator [this work] & \makecell{1.5 V at 18 GHz\\ 1.6 V at 25.5 GHz }  & \makecell{Recycled TFLN\\ dual pass} \\ [0pt]
         \hline
         TFLN phase modulator \cite{zhang2023power} & \makecell{2.5 V at 20 GHz\\ 2.0 V at 25 GHz }& \makecell{Recycled TFLN\\ quad pass}\\ [0pt]
         \hline
         TFLN phase modulator \cite{yu2022integrated} & \makecell{2.6 V at 18.5 GHz\\ 2.3 V at 21.5 GHz} & \makecell{Recycled TFLN\\ dual pass} \\ [0pt]
         \hline
         TFLN phase modulator \cite{ren2019integrated} & 4.1 V at 20 GHz    & Single pass TFLN \\ [0pt]
         \hline
         Commercial phase modulator \cite{EOSPACE} & 3.8 V at 18 GHz     & Single pass bulk LN\\ [0pt]
         \hline
         Commercial phase modulator \cite{ ThorlabsPM} & 4.0 V at 18 GHz  &  Single pass bulk LN \\
        \hline
    \end{tabular}}
    \caption{Summary of the V$_{\pi}$  and architecture of various LiNbO$_3$ (LN) phase modulators.}
    \label{tab2}
\end{table}

\section{Summary of various low V$_{\pi}$ phase modulators}

Table \ref{tab2} shows the summary of V$_{\pi}$ and architecture of various thin film LiNbO$_3$ phase modulators \cite{zhang2023power,yu2022integrated,ren2019integrated}, and commerical bulk LiNbO$_3$ phase modulators \cite{EOSPACE, ThorlabsPM}. The TFLN phase modulator chip developed in this work features a measured   V$_{\pi}$ of 1.5V at 18 GHz, and  V$_{\pi}$ of 1.6V at 25.5 GHz, which is the record-low V$_{\pi}$ for LiNbO$_3$ phase modulators at telecomm C-band wavelength. \textcolor{black}{Note that various low V$_{\pi}$ TFLN intensity modulators are not included in Table  \ref{tab2}, as intensity modulators are not used for broadband EO comb generation in this work. }

\end{document}